\documentclass[10pt]{amsart}
\usepackage{amsmath,amsfonts,amssymb,amsthm,amscd}
\newtheorem{thm}{Theorem}
\newtheorem{lem}[thm]{Lemma}
\newtheorem{defn}[thm]{Definition}
\newtheorem{prop}[thm]{Proposition}
\newtheorem{cor}[thm]{Corollary}
\newtheorem{example}[thm]{Example}

\def\res{\mathop{\rm res}\limits}
\newcommand{\dbyd}[1]{\frac{\partial~}{\partial #1}}
\begin{document}


\title[Logarithmic Deformations of WDVV solutions]{Logarithmic deformations of the rational superpotential/Landau-Ginzburg construction of
solutions of the WDVV equations}

\author{James T. Ferguson and Ian A. B. Strachan}

\date{30$^{\rm th}$ May, 2006}

\address{Department of Mathematics\\ University of Glasgow\\ Glasgow G12 8QQ\\ U.K.}

\email{j.ferguson@maths.gla.ac.uk, i.strachan@maths.gla.ac.uk}

\keywords{Frobenius manifolds, WDVV equations, dispersionless KP hierarchy}
\subjclass{11F55, 53B50, 53D45}

\begin{abstract}
The superpotential in the Landau-Ginzburg construction of solutions to the
Witten-Dijkgraaf-Verlinde-Verlinde (or WDVV) equations is modified
to include logarithmic terms. This results in
deformations - quadratic in the deformation parameters - of the normal
prepotential solution of the WDVV equations. Such solution satisfy various pseudo-quasi-homogeneity
conditions, on assigning a notional weight to the deformation parameters. This construction
includes, as a special case, deformations which are polynomial in the flat coordinates, resulting
in a new class of polynomial solutions of the WDVV equations.

\end{abstract}

\maketitle


\section{Introduction}

One of the most basic classes of Frobenius manifolds is comprised of those which are
defined on orbit spaces $\mathbb{C}^n/W\,,$ $W$ being a finite Coxeter group \cite{dubrovin1}. Following
from the observation of Arnold that the three polynomial solutions in 3-dimensions
were related to the Coxeter numbers of the Platonic solids it was realized that
the earlier Saito construction \cite{saito} provided a construction of Frobenius manifolds
and that the prepotentials (solutions to the WDVV-equations - see below) were
automatically polynomial with respect to a distinguished coordinate system,
the so-called flat coordinates $\{t^i\}\,.$

Such prepotentials are quasihomogeneous, a property that may be expressed
in terms of an Euler vector field
\[
E = \sum_i d^i t^i \dbyd{t^i}
\]
as
\[
\mathcal{L}_E F = (2h+2) F\,,
\]
where the $d^i$ are the degrees of the basic $W$-invariant polynomials
and $h$ is the Coxeter number of $W\,.$ Such solutions are semi-simple and it
was conjectured by Dubrovin that all semi-simple polynomial solutions
arise from this construction for some Coxeter group. This was later
proved by Hertling \cite{hert}.

In this paper we construct a new class of semi-simple polynomial solutions to the WDVV equations.
This does not contradict the result of Hertling as the solution does not satisfy the
full set of axioms of a Frobenius manifold, in particular the solutions are not quasi-homogeneous. These
solutions may be regarded as a deformation of the $A_N$-polynomial solutions, in the
sense that the prepotential takes the form
\[
F(t^1\,,\ldots\,,t^N\,,b) = F^{(0)}(t^1\,,\ldots\,,t^N)+ k F^{(1)}(t^1\,,\ldots\,,t^N\,,b)
\]
where $F^{(0)}$ is the polynomial solutions defining the Frobenius manifold structure on the
space $\mathbb{C}^N/A_N$ and $k$ is some deformation parameter. Such solutions satisfy
a
pseudo-quasi-homogeneity condition. With the Euler vector field
\[
E=\sum_{i=1}^N (N+2-i) t^i \dbyd{t^i} + b \dbyd{b}
\]
each part is separately quasi-homogeneous:
\begin{eqnarray*}
\mathcal{L}_E F^{(0)} & = & (2 N+4) F^{(0)}\,,\\
\mathcal{L}_E F^{(1)} & = & ( N+3) F^{(1)}\,.
\end{eqnarray*}
By assigning a fictitious scaling degree of $(N+1)$ to the deformation
parameter $k$ the full solution may thought of a pseudo-quasi-homogeneous.
These solutions will appear as a special case of a more general
construction.

The Frobenius manifold structure on the orbit space $\mathbb{C}^N/A_N$ may
also be derived \cite{dubrovin1, k1, k2} via a Landau-Ginzburg formalism as the
structure on the parameter space
of polynomials of the form
\begin{equation}
\lambda(p) = p^{N+1} + s_1 p^{N-1} + \ldots + s_N\,.
\label{ANpolynomial}
\end{equation}
More explicitly, the metric
\begin{equation}
\eta(\partial_{s_i},\partial_{s_j})  =   -\sum \res_{d\lambda=0}
\left\{ \frac{\partial_{s_i}\lambda(p) \,
\partial_{s_j}\lambda(p)}{\lambda'(p)}\,dp\right\} \label{metric}
\end{equation}
is flat (though, in these variables, it does not have constant
entries) and the tensor
\begin{equation}
c\,(\partial_{s_i},\partial_{s_j},\partial_{s_k})  = -\sum
\res_{d\lambda=0} \left\{ \frac{\partial_{s_i}\lambda(p) \,
\partial_{s_j}\lambda(p)\,
\partial_{s_k}\lambda(p)}{\lambda'(p)}\,dp\right\}
\label{multiplication}
\end{equation}
defines a totally symmetric $(3,0)$-tensor which further satisfies
various potentiality conditions from which one may construct a
so-called prepotential $F$ which satisfies the
Witten-Dijkgraaf-Verlinde-Verlinde (or WDVV) equations of
associativity
\[
\frac{\partial^3F}{\partial t^\alpha \partial t^\beta \partial t^\lambda}\eta^{\lambda\mu}
\frac{\partial^3F}{\partial t^\mu \partial t^\gamma \partial t^\delta}-
\frac{\partial^3F}{\partial t^\delta \partial t^\beta \partial t^\lambda}\eta^{\lambda\mu}
\frac{\partial^3F}{\partial t^\mu \partial t^\gamma \partial t^\alpha}=0
\,,\quad \alpha\,,\beta\,,\gamma\,,\delta=1\,\ldots\,,N
\]
where the coordinates $\{t^i\}$ are a set of flat coordinates for the metric $\eta$ defined by
(\ref{metric}).
Geometrically, a solution defines a multiplication $\circ: TM \times TM \rightarrow TM$ of vector fields on the
parameter space $M\,,$ i.e.
\begin{eqnarray*}
\partial_{t^\alpha} \circ \partial_{t^\beta} & = &\left(
\frac{\partial^3F}{\partial t^\alpha \partial t^\beta \partial t^\sigma} \eta^{\sigma \gamma}\right)
\partial_{t^\gamma}\,,\\
& := & c_{\alpha\beta}^\gamma(t)\,\partial_{t^\gamma}\,,
\end{eqnarray*}
the metric $\eta$ being used to raise and lower indices.

\begin{example}
With
\[
\lambda(p) = p^4 + s_1 p^2 + s_2 p + s_3
\]
the formula (\ref{metric}) gives the metric\footnote{In all examples indices are lowered for notational convenience}
\[
\eta = \frac{1}{2} ds_1 ds_3 + \frac{1}{4} ds_2^2 - \frac{s_1}{8} ds_1^2\,.
\]
While this metric is flat, the $s^i$ are not flat coordinates. With
\begin{eqnarray*}
s_3 & = & t_1 + \frac{1}{8} t_3^2\,,\\
s_2 & = & t_2 \,, \\
s_1 & = & t_3
\end{eqnarray*}
one obtains a metric with constant coefficients. The tensor given by
the formula (\ref{multiplication}) may then be used to construct the prepotential
\[
F=\frac{1}{8} t_1^2 t_3 + \frac{1}{8} t_1 t_2^2 - \frac{1}{64} t_2^2 t_3^2 + \frac{1}{3840} t_3^5\,.
\]
\end{example}

Such polynomial solution may be seen from a variety of different points of view
(and part of the rich mathematical structure of Frobenius manifold arises as
from the fact that it lies at the intersection of seemingly disconnected areas of mathematics):

\begin{itemize}

\item[(i)] as a basic example of an orbit space construction. Here the manifold
is $\mathbb{C}^n/A_N$ where $A_N$ is a Coxeter group;

\item[(ii)] as a topological Landau-Ginsburg field theory;

\item[(iii)] as a reduction of the dispersionless KP hierarchy.

\end{itemize}
The point of view that will be taken in this paper is last, i.e. that a solution
to the WDVV equations may be obtained from a specific reduction of the dispersionless
KP hierarchy \cite{k1, k2}. In particular it will be shown that the so-called water-bag reduction
of the KP hierarchy \cite{GibbonsT} (see also \cite{BogKon}) also results in a solutions of the WDVV equations, though not, as in earlier examples,
a full Frobenius manifold because of the non-existence of an Euler vector
field. This builds on a recent preprint \cite{chang} where a 2-component system was studied.

\section{The dispersionless KP hierarchy}

The dispersionless KP (or dKP) hierarchy is defined in terms of a Lax function
\[
\lambda(p) = p + \sum_{n=1}^\infty u_{n}(x,t) p^{-n}
\]
by the Lax equation
\[
\partial_{T_n} \lambda(p)= \left\{ \lambda(p) , \left[ \lambda^n(p)_{+} \right] \right\}
\]
where $\{f,g\}=f_x g_p - f_p g_x$ is the ordinary Poisson bracket and
$\left[\,\right]_{+}$ denotes the projection onto non-negative powers of $p\,.$ Various
reduction of this infinite component hierarchy have been studied, the most fundamental
being the $A_N$-reduction
\[
\lambda(p) = \left[p^{N+1} + s_1 p^{N-1} + \ldots + s_N\right]^\frac{1}{N+1}
\]
and this leads to a Frobenius manifold structure, defined above,
on the space of parameters $\{s_i\}\,.$ More recently a so-called
\lq water-bag\rq~ reduction has been studied, where one takes
\[
\lambda(p) = p + \sum_{i=1}^N k_i \log \left(
\frac{p-p_i}{p-{\tilde p}_i}\right)\,.
\]
In a recent preprint Chang \cite{chang} showed that in the $N=1$ case one may
construct a solution of the WDVV equation by analysing the
recursion relations satisfied by the conservation laws of the
associated 2-component dispersionless hierarchy. Here we
generalise this setting and consider functions of the form
\[
\lambda(p) = \left({\rm rational~function}\right)(p) +
\sum_{i=1}^M k_i \log(p-b_i)
\]
Formally one may
expand this function for large $p$ as a series, but this will have terms of the
form
\[
\left(\sum_{i=1}^M k_i\right) \log p\,
\]
and the constraint $\sum k_i=0$ is often imposed. Here we show
that one still gets a solution without such a constraint. To make
$\lambda$ single valued one has to make various cuts on the complex
plane. For simplicity we present proofs in the polynomial case,
with
\begin{equation}
\lambda(p) = p^{N+1} + s_1 p^{N-1} + \ldots + s_N+ \sum_{i=1}^M k_i
\log(p-b_i) \label{waterbag}
\end{equation}
and state the result for the rational case - no essential new
features will be present in the rational case that are not already
present in the polynomial case. Note that without this constraint
the function is not technically a reduction of the dKP hierarchy,
but one may associated a \lq regularised\rq~function
\[
\lambda(p) \rightarrow \lambda(p) -\left(\sum_{i=1}^M k_i\right)
\log p
\]
which is \cite{pavlov}. For this reason we call the form (\ref{waterbag}) a generalised
water-bag reduction. We denote the space of such superpotentials $\mathcal{M}^{(M,N)}$ or
just $\mathcal{M}\,.$

\section{Solutions of the WDVV equations from the generalised Water-bag reduction of the dispersionless KP hierarchy}

We begin by proving that the formulae
(\ref{metric},\ref{multiplication}) with the function
(\ref{waterbag}) define a commutative, associative, semi-simple
multiplication on the tangent space to the manifold of parameters.
This will be done using canonical coordinates - the critial values
of $\lambda$ (i.e. $\lambda$ evaluated at its critical points).
Since $\lambda(p)$ only involves logarithms its derivative is a
rational function which may be written in the form
\[ \lambda^{\prime}(p) = \frac{ (N+1)\prod_{i=1}^{M+N} (p -
\xi_i) }{\prod_{j=1}^M (p-b_j)}
\]
(we assume that we are considering the generic case, where the
poles and zeros are all distinct). The canonical coordinates are
then
\[
u^i = \lambda(\xi_i)\,,\qquad\qquad i=1\,\ldots\,,N+M\,
\]
(for such a formula to be single-valued, various cuts have to be made in the complex plane). The
proof follows \cite{dubrovin1}, Lemma 4.5. From the formulae
\[
\left.\frac{\partial~}{\partial u^i} \lambda(p) \right|_{p=\xi_j} = \delta_{ij}\,,\qquad\qquad i=1\,\ldots\,,N+M
\]
and
\[
\frac{\partial~}{\partial u^i} \lambda(p) = \left\{ \prod_{r=1}^M (p-b_r)\right\}^{-1} B_i(p)
\]
(where $B_i$ is a polynomial of degree $N+M-1$) one obtains
\[
B_i(\xi_j) = \begin{cases} 0\,, &  i\neq j\\ \prod_{r=1}^M (\xi_i-b_r)\,, &  i=j\,. \end{cases}
\]
The Lagrange interpolation formula then gives
\[
B_i(p) = \frac{ \prod_{j\neq i} (p-\xi_j) \prod_{r=1}^M (\xi_i-b_r)}{\prod_{j\neq i} (\xi_i-\xi_j)}
\]
and hence
\begin{eqnarray}
\frac{\partial\lambda(p)}{\partial u^i} & = &
\frac{ \prod_{j\neq i} (p-\xi_j) \prod_{r=1}^M (\xi_i-b_r)}{\prod_{j\neq i} (\xi_i-\xi_j) \prod_{r=1}^M (p-b_r)}\,,\nonumber\\
& = & \frac{1}{(p-\xi_i)} \lambda^{\prime}(p)
\left\{ \frac{\prod_{r=1}^M (\xi_i-b_r)}{\prod_{j\neq i} (\xi_i-\xi_j)} \right\}\,, \nonumber\\
& = & \frac{1}{(p-\xi_i)}
\frac{\lambda^{\prime}(p)}{\lambda^{\prime\prime}(\xi_i)}\,.\label{dlbydu}
\end{eqnarray}
Note that this is the same functional form as in the polynomial
case. With this
\begin{eqnarray*}
\eta(\partial_{u_i},\partial_{u_j}) & = & -\sum \res_{d\lambda=0}
\left\{ \frac{1}{(p-\xi_i)(p-\xi_j)} \frac{
\lambda^{\prime}(p)}{\lambda^{\prime\prime}(\xi_i)\lambda^{\prime\prime}(\xi_j)}dp
\right\}\,,\\
& = & - \frac{1}{\lambda^{\prime\prime}(\xi_i)} \delta_{ij}\,.
\end{eqnarray*}
Note that while log-terms appear in $\lambda\,,$ the metric
formula involves derivatives of $\lambda$ and hence involves
rational functions only.

Similarly
\[
c(\partial_{u_i},\partial_{u_j},\partial_{u_k}) =
\begin{cases}
\displaystyle{-\frac{1}{\lambda^{\prime\prime}(\xi_i)} }\,,& i=j=k\,,\\
\hskip 5mm 0\,, & {\rm otherwise\,.} \end{cases}
\]
Collecting these results one arrives at the following:
\begin{lem}\label{lem:compat}
The formulae (\ref{metric}) and (\ref{multiplication}) with $\lambda$ given by (\ref{waterbag}) define, at a generic point, a semi-simple, commutative, associative multiplication
\begin{equation}
\frac{\partial~}{\partial u^i} \circ \frac{\partial~}{\partial u^j} = \delta_{ij} \frac{\partial~}{\partial u_i}\,,
\label{semisimple}
\end{equation}
compatible with the metric
\begin{equation}
\eta=-\sum_{r=1}^{M+N} \frac{du_i^2}{\lambda^{\prime\prime}(\xi_i)}
\label{diagonalmetric}
\end{equation}
\end{lem}

\medskip

\noindent This multiplication has an identity. Since $e(\lambda)=1\,,$ where the vector field $e$
is defined to be
\[
e=\frac{\partial~}{\partial s^N}\,,
\]
it is immediate from equations (\ref{metric}) and (\ref{multiplication}) that
\[
c(\partial, \partial^\prime,e) = \eta(\partial,\partial^\prime)\,.
\]
From this it follows that $e$ is the identity for the multiplication. In semi-simple
coordinates it follows from the multiplication (\ref{semisimple}) that
\[
e=\sum_{r=1}^{M+N} \frac{\partial~}{\partial u^i}\,.
\]

\medskip

We prove next that the metric is flat and Ergoff. In the
pure-polynomial case (or $A_N$-case) the flat coordinates are
defined by an inverse series, using the so-called thermodynamic
identity. The presence of the logarithms makes such an inversion
problematical. However, it turns out that part of the
flat-coordinates of the metric are exactly the same as in the
polynomial case.

\begin{lem}\label{lemma:metric}
The formula (\ref{metric}) with $\lambda$ given by (\ref{waterbag}) gives the following:

\[
\eta(\partial_{s_i},\partial_{s_j}) = -\sum \res_{d\lambda_{+}=0}
\left\{ \frac{\partial_{s_i} \lambda_{+}(p) \partial_{s_j}
\lambda_{+}(p) }{\lambda_{+}^\prime(p)} dp \right\}\,,\qquad\qquad
i\,,j=1\,,\ldots\,, N\,,
\]
where $\lambda_{+}(p) =p^{N+1} + s_1 p^{N-1} + \ldots + s_N$ is a
truncation of $\lambda\,,$ and
\begin{eqnarray*}
\eta(\partial_{b_r},\partial_{s_j}) & = & 0\,,\qquad\qquad r=1\,,\ldots\,,M\,,j=1\,,\ldots\,,N\,,\\
\eta(\partial_{b_i},\partial_{b_j}) & = &k_i \delta_{ij}\,,
\qquad\qquad i,j=1\,,\ldots\,,M\,.
\end{eqnarray*}

\end{lem}

\noindent It follows from these formulae that the metric is flat.

\begin{proof} These formulae just involve the use of basic ideas from
complex variable theory.
\begin{eqnarray*}
\eta(\partial_{s_i},\partial_{s_j}) & = &-\sum \res_{d\lambda_=0}
\left\{
\frac{p^{2N-i-j} }{\lambda^\prime(p)}dp\right\}\,, \\
& = & \res_{p=\infty}\left\{ \frac{p^{2N-i-j}
}{\lambda^\prime(p)}dp\right\}\,.
\end{eqnarray*}

Now
\begin{eqnarray*}
\lambda^\prime(p) & = & \lambda^\prime_{+}(p) + \sum_{r=1}^M \frac{k_i}{(p-b_i)}\,, \\
& = &\lambda^\prime_{+}(p) \left\{ 1 +
\frac{1}{\lambda^\prime_{+}(p)}\sum_{r=1}^M
\frac{k_i}{(p-b_i)}\right\}\,.
\end{eqnarray*}
Hence
\begin{eqnarray*}
\eta(\partial_{s_i},\partial_{s_j}) & = & \res_{p=\infty} \left\{
\frac{p^{2n-i-j}}{\lambda_{+}^\prime(p)} \left[ 1 +
\frac{1}{\lambda^\prime_{+}(p)}\sum_{r=1}^M \frac{k_i}{(p-b_i)}
\right]^{-1} dp \right\}\,,\\
&=& -\res_{{\tilde p}=0} \left\{ \frac{ {\tilde p}^{i+j-N-2} }{
\mu( {\tilde p}) } \left[ 1 + \frac{ {\tilde p}^{N+1} }{ \mu(
{\tilde p}) } \sum_{r=1}^M \frac{k_i}{1-{\tilde p} b_i}
\right]^{-1} d{\tilde p} \right\}\,,\\
& = & - \res_{{\tilde p}=0} \left\{ \frac{ {\tilde p}^{i+j-N-2} }{
\mu( {\tilde p}) }d{\tilde p} \right\}\,,
\end{eqnarray*}
where ${\tilde p}=p^{-1}$ and $\lambda^\prime_{+}(p) = {\tilde
p}^{-N}  \mu(\tilde p)\,.$ Reversing the argument yields the
result.

\noindent Similarly,
\begin{eqnarray*}
\eta(\partial_{s_i},\partial_{b_r}) & = & \sum\res_{d\lambda=0}
\left\{
\frac{p^{N-i}}{\lambda^\prime(p)} \, \frac{k_r}{(p-b_r)} dp \right\} \,,\\
& = & -\frac{1}{N+1}\res_{p=\infty} \left\{ \frac{ k_r p^{N-i}
\prod_{r \neq i} (p-b_r)}{\prod_{j=1}^{M+N} (p- \xi_j)} dp
\right\} \,,\\
& = & \frac{1}{N+1}\res_{{\tilde p}=0} \left\{ k_r p^{i-1} \frac{
\prod_{r\neq i} (1- b_r {\tilde p})}{\prod_{j=1}^{M+N} (1- \xi_j
{\tilde p})}
dp \right\}\,,\\
& = & 0\,.
\end{eqnarray*}

\noindent Finally,
\[
\eta(\partial_{b_i},\partial_{b_j})  =  - \frac{1}{N+1}\sum
\res_{d\lambda=0} \left\{
\frac{k_i}{(p-b_i)}\,\frac{k_j}{(p-b_j)}\, \frac{
\prod_{r=1}^M(p-b_r)}{\prod_{k=1}^{M+N} (p-\xi_k)} dp\right\}\,.
\]

For $i\neq j$ this, on deforming the contour around the Riemann
sphere, gives zero: there is no pole at infinity, and the simple
poles cancel. For $i=j\,, $
\begin{eqnarray*}
\eta(\partial_{b_i},\partial_{b_i}) & = & -k_i^2
\sum\res_{d\lambda=0} \left\{
\frac{1}{(p-b_i)^2} \frac{1}{\lambda^{\prime}(p)} dp\right\}\,,\\
& = & k_i^2 \frac{1}{N+1}\frac{\prod_{k\neq i} (b_i -
b_k)}{\prod_i (b_i - \xi_i)}\,.
\end{eqnarray*}

On evaluating the residue at the poles using the two different
formulae for $\lambda^{\prime}(p)\,,$
\[
(N+1) p^N + (N-1) s_1 p^{N-2} + \ldots s_1 + \sum_{r=1}^M
\frac{k_i}{(p-b_r)} = (N+1) \frac{ \prod_{i=1}^{M+N} (p - \xi_i)
}{\prod_{j=1}^M (p-b_j)}
\]
one obtains
\[
k_i =(N+1) \frac{\prod_i (b_i - \xi_i)}{\prod_{k\neq i} (b_i -
b_k)}
\]
from which the final formulae follows.
\end{proof}
 \begin{proof}
 (\lq Thermodynamical identity\rq~-type proof of flat coordinates)

 Following the polynomial case in \cite{dubrovin1}, invert $\lambda_+(p)$ as $$p_+(k) = k +
 \frac{1}{N+1} \left( \frac{t^N}{k} + \frac{t^{N-1}}{k^2} + \ldots
 +\frac{t^1}{k^{N}}\right) + O\left(\frac{1}{k^{N+1}}\right)\, ,$$
 where $\lambda_+=k^{N+1}$. Then
 \begin{eqnarray*}
 \lambda(p_+(k,t),t,b) &=&
 \lambda_+(p_+(k,t),t)+\sum_{i=1}^{M}k_i\log(p_+-b_i)\, ,\\
 &=& k^{N+1}+\sum_{i=1}^{M}k_i\log(p_+-b_i)\, .
 \end{eqnarray*}
 Differentiating with respect to $t^\alpha$ gives
 \begin{eqnarray*}
 \left.\frac{d\lambda}{dp}\right|_{p=p_+(k)}\frac{\partial p_+}{\partial
 t^\alpha}+
 \frac{\partial\lambda}{\partial t^\alpha} &=&
 \sum_{i=1}^{M}\frac{k_i}{p_+-b_i}\frac{\partial p_+}{\partial
 t^\alpha}\, .\\
 &=& O\left(\frac{1}{k^{N+2-\alpha}}\right)\, .
 \end{eqnarray*}
 So we have as our thermodynamical identity in this case
 $$\dbyd{t^\alpha}(\lambda dp)+\dbyd{t^\alpha}(p_+d\lambda) =
 O\left(\frac{1}{k^{N+1-\alpha}}\right)dk\, .$$
 Although the right hand side is not zero as it is for
 polynomial $\lambda$, this identity is sufficient to give
 $$\dbyd{t^\alpha}(\lambda dp) = -k^{\alpha-1}dk+O\left(\frac{1}{k}\right)dk$$
 (eqn. (4.68) in \cite{dubrovin1}), from which it
 follows, using
 \[
 d\lambda = d\lambda_{+} + O\left(\frac{1}{k}\right) \,dk\,,
 \]
 that
 $$\eta(\partial_{t^\alpha},\partial_{t^\beta})=-\frac{\delta_{\alpha+\beta,N+1}}{N+1}\, .$$
 \end{proof}

The flat coordinates are therefore
\[
\{ t^i\,, i=1\,,\ldots\,,N\,;
b_j\,,j=1\,,\ldots\,,M\}
\]
where the $t^i$ are defined by the inverse series for the
truncated function $\lambda_{+}=\lambda_{+}(p)$, expanded as a
Puiseaux series as $\lambda\rightarrow \infty\,,$
 \begin{equation}\label{puiseaux}
  p(k) = k +
\frac{1}{N+1} \left( \frac{t^N}{k} + \frac{t^{N-1}}{k^2} + \ldots
+\frac{t^1}{k^{N}}\right) + O\left(\frac{1}{k^{N+1}}\right)
 \end{equation}
 where
$k=(\lambda_{+})^{\frac{1}{N+1}}\,,$ in the standard way
\cite{dubrovin1}. Note that each $t^i$ is a polynomial in the
$s_i$ and vice versa.

\medskip

Consider the diagonal metric (\ref{diagonalmetric}). Its rotation coefficients $\beta_{ij}$ are defined
by the formula
\[
\beta_{ij} = \frac{\partial_{u_i} H_j}{H_i}\,,\qquad H_i^2 = \frac{1}{\lambda^{\prime\prime}(\xi_i)}\,.
\]
Such a metric is said to be Egoroff if the rotation coefficients are symmetric. This then implies
that the metric may be written in terms of a single potential function $V(u)\,,$
\[
\eta=\sum_{i=1}^{M+N} \frac{\partial V}{\partial u^i} \left(du^i\right)^2\,.
\]
\begin{lem}
The metric (\ref{diagonalmetric}) is Egoroff.
\end{lem}
\begin{proof}
 In canonical coordinates $\eta$ is diagonal with $i^{th}$ entry
 $$-\frac{1}{\lambda''(\xi_i)}\, .$$

 From (\ref{dlbydu})
 \begin{eqnarray*}
 \frac{\partial\lambda}{\partial
 u^i} &=& \frac{1}{p-\xi_i}\frac{\lambda'(p)}{\lambda''(\xi_i)}\, ,\\
 &=& \frac{N+1}{\lambda''(\xi_i)}\frac{\prod_{r\neq
 i}(p-\xi_r)}{\prod_{s=1}^{M}(p-b_s)}\, ,
 \end{eqnarray*}
 so we have
 $$\frac{\partial\lambda}{\partial
 u^i}\prod_{s=1}^{m}(p-b_s)=\frac{N+1}{\lambda''(\xi_i)}\prod_{r\neq i}(p-\xi_i)$$
 where each side is a polynomial of degree $N+M-1\,.$

 Also
 $$\frac{\partial\lambda}{\partial u^i}=\frac{\partial
 s_1}{\partial u^i}p^{N-1}+\frac{\partial
 s_2}{\partial u^i}p^{N-2}+\dots+\frac{\partial
 s_N}{\partial u^i}-\sum_{r=1}^{M}\frac{k_r}{p-b_r}\frac{\partial
 b_r}{\partial u^i}\, ,$$
 so
 $$\frac{\partial\lambda}{\partial u^i}\prod_{s=1}^{m}(p-b_s)=\left(\frac{\partial
 s_1}{\partial u^i}p^{N-1}+\dots+\frac{\partial
 s_N}{\partial u^i}\right)\prod_{s=1}^{M}(p-b_s)-\sum_{r=1}^{M}k_r\frac{\partial
 b_r}{\partial u^i}\prod_{s\neq r}(p-b_s)\, .$$

 Comparing coefficients of $p^{N+M-1}$ in $\frac{\partial\lambda}{\partial
 u^i}\prod_{s=1}^{M}(p-b_s)$ in these two expressions gives
 $$\frac{N+1}{\lambda''(\xi_i)}=\frac{\partial s_1}{\partial u^i}\, .$$

 Hence
 $$\eta(\frac{\partial}{\partial u^i},\frac{\partial}{\partial u^i})=-\frac{1}{\lambda''(\xi_i)}=\frac{\partial}{\partial u^i}\left(-\frac{1}{N+1}s_1\right)\,.$$
\end{proof}
This Egoroff property is equivalent to a potentiality condition on
the $(3,0)$-tensor $c\,,$ namely that the tensor $\nabla c$ is
totally symmetric. Since the metric is flat one may, in
flat-coordinates, integrate by Poincar\'e's lemma and express
everything in terms of a prepotential $F$ which satisfies the WDVV
equations. Collecting these results together one obtains:
\begin{prop}
The flat metric (\ref{metric}) and totally symmetric $(3,0)$ tensor (\ref{multiplication}), with $\lambda$
given by
\[
\lambda = p^{N+1} + s_1 p^{N-1} + \ldots + s_N+ \sum_{i=1}^M k_i
\log(p-b_i)\,,\quad k_i {\rm~constant}
\]
define, on the space $\mathcal{M}^{(M,N)}$ a solution to the WDVV
equations. Geometrically they define a semi-simple, associative,
commutative algebra with unity on the tangent space $T\mathcal{M}$
compatible with the flat metric.
\end{prop}

\medskip

Before giving some examples, it must be remarked that we do not
have a Frobenius manifold, just a solution to the WDVV equations.
As was remarked in one of the earliest papers on water-bag
reductions, such reductions do have have a scaling symmetry and
this fact manifests itself in the non-existence of an Euler vector
field, the existence of which is part of the definition of a
Frobenius manifold (though it should be remarked that some authors
do not require such a field in their definition, denoting
manifolds with such a field as a conformal Frobenius manifold).

\begin{example} $N=0\,,M=2\,.$
In the above proofs it has been assumed that $N\neq 0\,.$ However one
may adapt these proofs to deal with this case. In particular, the
identity field, normally associated to the variable $s_N$, has to
be carefully defined. With
\[
\lambda(p) = p + k_1 \log\left[p-(t_1+t_2)\right] + k_2
\log\left[p-(t_1-t_2)\right]
\]
one obtains the prepotential
\[
F=\frac{1}{6} \left\{ k_1 (t_1+t_2)^3 + k_2 (t_1-t_2)^3 \right\} +
2 k_1 k_2 \,t_2^2 \,\log t_2\,.
\]
\end{example}
\noindent Note that if the condition $k_1+k_2=0$ is imposed, one obtains,
after some rescalings, the solution obtained by Chang. This
example was the original motivation of this work.

 \begin{lem}\label{lemma:multiplication}
 \begin{eqnarray*}
 c\left(\dbyd{b_\alpha},\dbyd{b_\beta},\dbyd{b_\gamma}\right)&=&0\,,\hspace{1cm}\alpha,\beta,\gamma\text{
 distinct\,,}\\
 c\left(\dbyd{b_\alpha},\dbyd{b_\alpha},\dbyd{b_\beta}\right)&=&\frac{k_\alpha
 k_\beta}{b_\beta-b_\alpha}\,,\hspace{1cm}\alpha\neq\beta\,,\\
 c\left(\dbyd{b_\alpha},\dbyd{b_\alpha},\dbyd{b_\alpha}\right)&=&k_\alpha\lambda'_+(b_\alpha)+\sum_{r\neq\alpha}\frac{k_\alpha
 k_r}{b_\alpha-b_r}\,,\\
&&\\
&&\\
 c\left(\dbyd{b_\alpha},\dbyd{b_\beta},\dbyd{s_\gamma}\right)&=&0\,,\hspace{1cm}\alpha\neq\beta\,,\\
 c\left(\dbyd{b_\alpha},\dbyd{b_\alpha},\dbyd{s_\gamma}\right)&=&k_\alpha(b_\alpha)^{N-\gamma}\,,\\
 &&\\
 &&\\
c\left(\dbyd{b_\alpha},\dbyd{s_\beta},\dbyd{s_\gamma}\right)&=&k_\alpha
 S_{\beta+\gamma}(s_1,\dots,s_N,b_\alpha)\,,
 \end{eqnarray*}

 $$c\left(\dbyd{s_\alpha},\dbyd{s_\beta},\dbyd{s_\gamma}\right)=R_{\alpha+\beta+\gamma}^{(0)}(s_1,\dots,s_N)+\sum_{j=1}^{M}k_jR_{\alpha+\beta+\gamma}^{(1)}(s_1,\dots,s_N,b_j)$$
 where $S_\sigma$, $R_\sigma^{(0)}$ and $R_\sigma^{(1)}$ are polynomial functions of their respective variables,
 and independent of all $k_i$'s.

 In particular, the term independent of $k_j$,
 $R_{\alpha+\beta+\gamma}^{(0)}(s_1,\dots,s_N)$,
 is precisely the value of
 $c(\partial_{s_\alpha},\partial_{s_\beta},\partial_{s_\gamma})$
 found from (\ref{multiplication}) using the polynomial $\lambda_+(p)$
 as the Landau-Ginzburg potential (\ref{ANpolynomial}).

 \begin{proof}
 Here we write $$\lambda'(p)=\frac{\nu(p)}{\prod_{j=1}^{M}(p-b_j)}$$
 where
 \begin{eqnarray*}
 \nu(p)&=&\lambda'_+(p)\prod_{j=1}^{M}(p-b_j)+\sum_{j=1}^{M}k_j\prod_{k\neq
 j}(p-b_k)\, ,\\
 &=&(N+1)\prod_{j=1}^{M}(p-\xi_j)\, .
 \end{eqnarray*}
 After the substitution $p\rightarrow 1/{\tilde p}$ we will have cause to
 refer to the polynomial
 $$\mu({\tilde p})={\tilde p}^N\lambda'_+\left(\frac{1}{\tilde p}\right)=(N+1)+(N-1)s_1{\tilde p}^2+(N-2)s_2{\tilde p}^3+\dots+s_{N-1}{\tilde
 p}^N\, .$$

 \bigskip\noindent({\bf{bbb}}) From the definition (\ref{multiplication}),
 $$c\left(\dbyd{b_\alpha},\dbyd{b_\beta},\dbyd{b_\gamma}\right)=\sum\res_{\nu=0}\frac{k_\alpha
 k_\beta
 k_\gamma}{(p-b_\alpha)(p-b_\beta)(p-b_\gamma)}\frac{\prod_{j=1}^M(p-b_j)}{\nu(p)}dp\, .$$
 This is evaluated by deforming the contour to encompass the poles
 at $p=\infty$ and possibly at $p=b_\alpha$ if there is
 repetition in the $b$'s. The residue at infinity is zero, and so
 in particular
 $c(\partial_{b_\alpha},\partial_{b_\beta},\partial_{b_\gamma})=0$
 for $\alpha,\beta,\gamma$ distinct.

 For the case $(\alpha,\alpha,\beta)$, the pole at $p=b_\alpha$ is
 simple, and the result follows immediately, noting that
 $\nu(b_\alpha)=k_\alpha\prod_{k\neq\alpha}(b_\alpha-b_k)$.

 For the case $\alpha=\beta=\gamma$, the pole is second order, and
 is evaluated directly as
 \begin{eqnarray*}
 c\left(\dbyd{b_\alpha},\dbyd{b_\alpha},\dbyd{b_\alpha}\right)&=&-\res_{p=b_\alpha}\frac{k_\alpha^3}{(p-b_\alpha)^2}\frac{\prod_{k\neq\alpha}(p-b_k)}{\nu(p)}dp\, ,\\
 &=&-k_\alpha^3\left.\frac{d}{dp}\right|_{p=b_\alpha}\frac{\prod_{k\neq\alpha}(p-b_k)}{\nu(p)}\,.
 \end{eqnarray*}

 \bigskip\noindent{\bf{(bbs)}}
 \begin{eqnarray*}
 c\left(\dbyd{b_\alpha},\dbyd{b_\beta},\dbyd{s_\gamma}\right)&=&-\sum\res_{\nu=0}\frac{k_\alpha
 k_\beta}{(p-b_\alpha)(p-b_\beta)}\frac{p^{N-\gamma}\prod_{j=1}^{M}(p-b_j)}{\nu(p)}dp\,,\\
 &=&\left(\res_{p=\infty}+\res_{p=b_\alpha}+\res_{p=b_\beta}\right)\frac{k_\alpha
 k_\beta}{(p-b_\alpha)(p-b_\beta)}\frac{p^{N-\gamma}\prod_{j=1}^{M}(p-b_j)}{\nu(p)}dp\,.
 \end{eqnarray*}

 Once again there is no pole at infinity, and there exists a (simple) pole at
 $p=b_\alpha$ only if $\alpha=\beta$. The result
 again follows from
 $\nu(b_\alpha)=k_\alpha\prod_{j\neq\alpha}(b_\alpha-b_j)$.

 \bigskip\noindent{\bf{(sss)}}
 \begin{eqnarray*}
 c\left(\dbyd{s_\alpha},\dbyd{s_\beta},\dbyd{s_\gamma}\right)&=&
 \res_{p=\infty}\frac{p^{3N-\alpha-\beta-\gamma}\prod_{j=1}^{M}(p-b_j)}
 {\lambda'_+(p)\prod_{j=1}^{M}(p-b_j)+\sum_{j=1}^{M}k_j\prod_{k\neq
 j}(p-b_k)}dp\,,\\
 &=&\res_{p=\infty}\frac{p^{3N-\alpha-\beta-\gamma}}{\lambda'_+(p)}
 \left[1+\sum_{j=1}^{M}\frac{k_j}{\lambda'_+(p)(p-b_j)}\right]^{-1}dp\,.
 \end{eqnarray*}
 This is expanded as a Taylor series in $x=\sum
 k_j/\lambda'_+(p)(p-b_j)$ to give a series of terms
 $$c\left(\dbyd{s_\alpha},\dbyd{s_\beta},\dbyd{s_\gamma}\right)=\sum_{i=0}^{\infty}\tilde{R}_{\alpha+\beta+\gamma}^{(i)}$$
 where
 $$\tilde{R}_{\sigma}^{(i)}=(-1)^{i+1}\res_{p=\infty}\frac{p^{3N-\sigma}}{\lambda'_+(p)}\left[\frac{1}{\lambda'_+(p)}\sum_{j=1}^{M}\frac{k_j}{p-b_j}\right]^idp\,.$$
 So, in particular,
 $R_{\alpha+\beta+\gamma}^{(0)}:=\tilde{R}_{\alpha+\beta+\gamma}^{(0)}=\res_{p=\infty}\frac{\partial_{s_\alpha}\lambda_+\partial_{s_\beta}\lambda_+\partial_{s_\gamma}\lambda_+}{\lambda'_+}dp$
 is $c_{\alpha\beta\gamma}$ from the $A_N$ orbit space corresponding to $\lambda_+$.

 $\tilde{R}_{\sigma}^{(1)}(s_1,\dots,s_N,b_1,\dots,b_M)$ can be decomposed as
 $\sum_{i=1}^{M}k_iR_{\sigma}^{(1)}(s_1,\dots,s_N,b_i)$ where
 \begin{eqnarray*}
 R_{\sigma}^{(1)}(s_1,\dots,s_N,b)&=&-\res_{p=\infty}\frac{p^{3N-\sigma}}{(p-b)(\lambda'_+(p))^2}dp\,,\\
 &=&\res_{\tilde{p}=0}\frac{1}{(1-b\tilde{p})(\mu(\tilde{p}))^2}\tilde{p}^{\sigma-N-1}d\tilde{p}\,.
 \end{eqnarray*}
 This is seen to be zero for $\sigma\geq N+1$, and $1/(N+1)^2$ for
 $\sigma=N$. For $\sigma<N$ it is a pole of order $N+1-\sigma$ and can
 be evaluated as
 \begin{equation}\label{diffresR1}
 \frac{1}{(N-\sigma)!}\left.\left(\frac{d}{d\tilde{p}}\right)^{N-\sigma}\right|_{\tilde{p}=0}\frac{1}{(1-b\tilde{p})(\mu(\tilde{p}))^2}\,.
 \end{equation}
 Clearly this evaluates to a polynomial in $\{s_1,\dots,s_N,b\}$.
 Finally, by making the substitution $p\rightarrow 1/\tilde{p}$ it
 can be seen that $\tilde{R}_\sigma^{(i)}=0$ for $i\geq 2$.

 \bigskip\noindent{\bf{(bss)}}
 Proceeding as in the (sss) case, we are led to
 $$c\left(\dbyd{b_\alpha},\dbyd{s_\beta},\dbyd{s_\gamma}\right)=k_\alpha\sum_{i=1}^{M}S_{\beta+\gamma}^{(i)}$$
 where
 \begin{eqnarray*}
 S_{\sigma}^{(i)}&=&(-1)^{i+1}\res_{p=\infty}\frac{p^{2N-\sigma}}{p-b_\alpha}\frac{1}{(\lambda'_+(p))^{i+1}}\left[\sum_{j=1}^{M}\frac{c_j}{p-b_j}\right]^{i}dp\,,\\
 &=&(-1)^i\res_{\tilde{p}=0}\frac{\tilde{p}^{\sigma-N-1+i(N+1)}}{(1-b_\alpha\tilde{p})(\mu(\tilde{p}))^{i+1}}\left[\frac{c_j}{1-b_j\tilde{p}}\right]^{i}d\tilde{p}\,.
 \end{eqnarray*}
 From this we can see that $S_\sigma^{(i)}=0$ for $i\geq 1$. This
 leaves only
 $$S_\sigma:=S_\sigma^{(0)}=\res_{\tilde{p}=0}\frac{\tilde{p}^{\sigma-N-1}}{(1-b_\alpha\tilde{p})\mu(\tilde{p})}d\tilde{p}\,,$$
 which is zero for $\sigma\geq N+1$, and $1/(N-1)$ for $\sigma=N$,
 whilst for $\sigma\leq N-1$ we evaluate as
 \begin{equation}\label{diffresS}
 \frac{1}{(N-\sigma)!}\left.\left(\frac{d}{d\tilde{p}}\right)^{N-\sigma}\right|_{\tilde{p}=0}\frac{1}{(1-b_\alpha\tilde{p})\mu(\tilde{p})}\,.
 \end{equation}

 \end{proof}
 \end{lem}

 For the Frobenius structure on the space of polynomials
 $$\lambda(p) = p^{N+1} + s_1 p^{N-1} + \ldots + s_N,$$ the
 variables $s_i$ inherit a scaling symmetry from the scaling of
 the polynomial. Namely if $p\rightarrow\epsilon p$ and we ask
 $\lambda\rightarrow\epsilon^{N+1}\lambda$, then we require
 $s_i\rightarrow\epsilon^{i+1}s_i$. Thus we conclude $s_i$ has
 degree $i+1$.

 For the water-bag reduction
 $$\lambda(p)=p^{N+1}+s_1p^{N-1}+\dots+s_N+\sum_{i=1}^{M} k_i\log(p-b_i),$$
 the same degrees may be attached to the coefficients $\{s_i\}$,
 and to preserve homogeneity of the arguments of the logarithms,
 each $b_i$ is assigned degree 1. If, in addition, an
 non-geometrically justified degree of $N+1$ is assigned to each $k_i$, then
 the regularised function $\lambda(p)-\sum k_i\log p$ is
 homogeneous of degree $N+1$.

 \begin{lem}
 Under the rescalings
 \begin{eqnarray*}
 s_i\:\rightarrow&\epsilon^{i+1}s_i & i=1\dots N\, ,\\
 b_i\:\rightarrow&\epsilon b_i & i=1\dots M\, ,\\
 k_i\:\rightarrow&\epsilon^{N+1}k_i & i=1\dots M
 \end{eqnarray*}
 the prepotential $F$ associated to the water-bag reduction(\ref{waterbag}) is
 homogeneous of degree $2N+4$.
 \label{homogeneity}
 \begin{proof}
 This may be verified from the explicit expressions for the
 components of the tensor $c(\partial,\partial',\partial'')$
 obtained in Lemma \ref{lemma:multiplication},
 remembering to add the degrees lost from differentiating along
 $\partial,\partial',\partial''$.

 In particular, for $c(\partial_{b_\alpha},\partial_{b_\alpha},\partial_{b_\alpha})=k_\alpha\lambda'_+(b_\alpha)+\sum_{r\neq\alpha}\frac{k_\alpha
 k_r}{b_\alpha-b_r}$ we note that $\lambda'_+(b_\alpha)=(N+1)(b_\alpha)^N+(N-1)s_1(b_\alpha)^{N-1}+\dots+s_{N-1}$ has
 degree $N$.

 The degrees of the polynomials $R_\sigma^{(0)}$, $R_\sigma^{(1)}$ and
 $S_\sigma$, when they are not zero or constant, can de determined
 from the differential expressions (\ref{diffresR1}),
 (\ref{diffresS}) and the corresponding expression for
 $R_\sigma^{(0)}$, which is
 $$R_\sigma^{(0)}=\left\{
  \begin{array}{cl}
  0&\sigma\geq 2N+2\\
  -1/(N+1)&\sigma=2N+1\\
  \frac{1}{(2N+1-\sigma)!}\left.\left(\frac{d}{d\tilde{p}}\right)^{2N+1-\sigma}\right|_{\tilde{p}=0}\frac{1}{\mu(\tilde{p})}&\sigma\leq
  2N
  \end{array}\right.\,.$$

 In this the degree of zero is undetermined, whilst for the middle case,
 the degree of a constant is 0. Integrating with respect to
 $s_\alpha$,$s_\beta$ and $s_\gamma$ adds to this degree
 $(\alpha+1)+(\beta+1)+(\gamma+1)=\sigma+3=2N+4$.
 In the final case, if $\tilde{p}=1/p$ is considered to have
 degree $-1$, then $\mu(\tilde{p})$ has degree zero. Thus on
 differentiation we obtain the quotient of two homogeneous
 polynomials with relative degrees $2N+1-\sigma$. Evaluation at
 $\tilde{p}=0$ merely makes this the ratio of constant terms, so
 that $R_\sigma^{(0)}$ has degree $2N+1-\sigma$. Integrating will add
 $\sigma+3$ to this, making $2N+4$ as required. $S_\sigma$ and
 $R_\sigma^{(1)}$ proceed similarly.
 \end{proof}
 \end{lem}

 The degrees of the flat coordinates $\{t^i, i=1\dots N\}$ are
 inherited from the polynomial transformations rules relating them to the $s_i$. They can also be
 deduced from the Puiseaux series (\ref{puiseaux}), in which we
 require both $p$ and $k$ to scale with degree 1, so that the
 degree of $t^i$ is $N+2-i$.

 We now draw together some simple observations, which follow
 immediately from lemmas \ref{lemma:metric},
 \ref{lemma:multiplication} and \ref{homogeneity}.

 \begin{prop}\label{cannotthinkofone}
 The prepotential is at most quadratic in the parameters $k_i\,,$
 that is, up to quadratic terms in the flat coordinates:
\begin{eqnarray*}
F(t^1,\dots,t^N,b^1,\dots,b^M) &=& F^{(0)}(t^1,\dots,t^N)\\&& +
\sum_i k_i F^{(1)}(t^1,\dots,t^N,b^i)\\&& + \sum_{i\neq j} k_i k_j
F^{(2)}(b^i,b^j)
\end{eqnarray*}
where $F^{(0)}\,,F^{(1)}\,,F^{(2)}$ are independent of the
parameters $k_i$. $F^{(0)}$ is the prepotential for the
$\mathbb{C}^N/A_N$ orbit space with $\lambda_+$ as the
Landau-Ginzburg potential, and as such is a polynomial in the flat
coordinates $\{t^1,\dots,t^N\}$. $F^{(1)}$ is also a polynomial,
and $$F^{(2)}(b^i,b^j)=\frac{1}{8}(b^i-b^j)^2\log(b^i-b^j)^2\,.$$
In place of quasi-homogeneity we have
\begin{eqnarray*}
{\rm deg}\left(F^{(0)}\right) & = & 2N + 4\,,\\
{\rm deg}\left(F^{(1)}\right) & = & N + 3\,,\\
{\rm deg}\left(F^{(2)}\right) & = & 2\,, \qquad{{\sl (modulo~quadratic~terms)}}\,.\\
\end{eqnarray*}
The structure functions for the Frobenius algebra are always at
most linear in the parameters $k_i\,,$ that is:
\[
{c_{\alpha\beta}}^{\gamma} = {c_{\alpha\beta}^{(0)}}^\gamma +
\sum_i k_i \,{c_{\alpha\beta}^{(i)}}^\gamma\,.
\]
where the $c_{\alpha\beta}^{(0)~\gamma}$ and
$c_{\alpha\beta}^{(i)~\gamma}$ are independent of the parameters.
\end{prop}

An important class of solutions are polynomial in the flat
coordinates.

 \begin{cor}
 For $M=1$, the prepotential on the space of functions
 $$\lambda(p)=p^{N+1}+s_1p^{N-1}+\dots+s_N+k \log(p-b)$$
 is polynomial in the flat coordinates $\{t^i,b\}$
 Conversely, if the prepotential is polynomial
 in the flat coordinates then $M=1\,$ (or $M=0$)\,.

 \end{cor}
 \begin{proof}
 This is an immediate consequence of the decomposition of $F$ given in
 Proposition \ref{cannotthinkofone}: the component $F^{(2)}$ contains
 all non-polynomial terms appearing in $F$, and is present if and
 only if $M\geq 2$.

%

 \end{proof}

We finish this main section with two simple examples.

\medskip

\begin{example}{~~~~~~~~~~}

\medskip

$\bullet$ $\bf N=2\,,M=1\,.$

\medskip
\noindent With
\[
\lambda(p) = p^3 + t_2 p + t_1 + k \log(p-t_3)
\]
one obtains the prepotential
\[
F = \frac{1}{6} t_1^2 t_2 - \frac{1}{2}k\, t_1 t_3^2 -
\frac{1}{216} t_2^4 - \frac{1}{6} k\,(t_2^2 t_3 + t_2 t_3^3) -
\frac{1}{20}k\, t_3^5\,.
\]

$\bullet$ {\bf $\bf N=1\,,M$ arbitrary.}

\medskip
\noindent In this case one has
\[
\lambda(p)=p^2 + t_1 + \sum_{i=1}^M k_i \log(p-b_i)\,.
\]
With this, lemmas \ref{lemma:metric} and
\ref{lemma:multiplication} give, on integrating, the following
prepotential:

\[
F=-\frac{1}{12} t_1^3 + \sum_{i=1}^M k_i \left\{ \frac{t_1
b_i^2}{2} + \frac{b_i^4}{12}\right\} + \frac{1}{8} \sum_{i\neq j}
k_i k_j (b_i-b_j)^2 \log (b_i-b_j)^2\,.
\]

\end{example}

\noindent We note that the $z^2\log z$-type terms have appeared in the
WDVV-literature before (see, for example, \cite{FV, MH}) but one normally
considers these are being derived as examples of dual Frobenius manifolds
\cite{dubrovin2}. Their functional form suggests the type of term that may
be present in a construction of deformed solutions for other Coxeter group orbit
spaces.

\section{Geometric and Algebraic Properties}

In this section we study certain geometric and algebraic
properties of the manifold.

\subsection{Geometric Properties}

An important addition structure on a Frobenius manifold is an addition
flat metric known as the intersection form, It plays a vital role in the
understanding of various properties of the manifold, such as the monodromy
properties of the Gauss-Manin connection and associated bi-Hamiltonian structures.
Following this, we defined a second metric on manifold; while this is not
flat, it shares many properties of the intersection form of a genuine Frobenius
manifold.

Before this, we normalise the Euler vector field, so
\begin{equation}
E=\frac{1}{N+1} \sum_{i=1}^N (N+2-i) t^i \frac{\partial~}{\partial t^i} + \frac{1}{N+1}
\sum_{j=1}^M b^j \frac{\partial~}{\partial b^j}\,.
\label{euler}
\end{equation}
\begin{defn} The metric $g$ on $\mathcal{M}$ is defined as:
\[
g^{-1}(\omega_1,\omega_2) = i_E (\omega_1,\omega_2)\,.
\]
\end{defn}
\noindent It follows immediately from this that
\[
g(E\circ u,v) = \eta(u,v)
\]
and, in components,
\[
g^{ij} = c^{ij}_k E^k\,.
\]
To understand the scaling properties of this metric we introduce
an extended Lie derivative $\mathcal{L}^{ext}_X\,,$
\[
\mathcal{L}^{ext}_X=\mathcal{L}_X +  \sum_{r=1}^M k^r \dbyd{k^r}\,,
\]
so, for an arbitrary tensor $\omega^{i\ldots j}_{a\ldots b}\,,$
\[
\left(\mathcal{L}^{ext}_X \omega \right)^{i\ldots j}_{a\ldots b} =
\left(\mathcal{L}_X \omega \right)^{i\ldots j}_{a\ldots b} + \sum_{r=1}^M k^r \dbyd{k^r}
\omega^{i\ldots j}_{a\ldots b}\,.
\]
This may be used to clarify the pseudo-quasi-homogeneity properties of the various
structures, for example
\[
\mathcal{L}^{ext}_E\,F = (3-d) F\,, \qquad\qquad d=\frac{N-1}{N+1}\,.
\]
Similarly the metrics $g$ and $\eta$ have various pseudo-quasi-homogeneity properties:
\begin{lem}
The following equations hold:
\[
\begin{array}{rclcrcl}
[e,E]&=&e\,,& & & &\\
&&&&&&\\
\mathcal{L}^{ext}_E g^{-1}&=& (d-1) g^{-1}\,,& & \mathcal{L}^{ext}_E\eta^{-1}&=& (d-2) \eta^{-1}\,,\\
&&&&&&\\
\mathcal{L}^{ext}_e g^{-1}&=& \eta^{-1}\,,& & \mathcal{L}^{ext}_e\eta^{-1}&=& 0\,.
\end{array}
\]
\end{lem}
However, the metric $g$ is not flat, and moreover, despite being linear in $t^1$ the
pencil $g_\Lambda^{-1} = g^{-1} + \Lambda \eta^{-1}$ does not define an almost compatible
pencil (the tensor $E\circ:T\mathcal{M}\rightarrow T\mathcal{M}$ fails to satisfy the
Nijenhuis condition \cite{DS}), let alone a compatible pencil. The role of this second metric is therefore
unclear. Given the origin of these structures in reductions of the dKP hierarchy one would
expect bi-Hamiltonian structures of differential-geometric type. One possibility is the metric
\[
g=\sum \frac{1}{u_i \lambda^{\prime\prime}(\xi_i)} du_i^2\,.
\]
This does define a non-local bi-Hamiltonian structure \cite{iabs} but finding its form in the flat-coordinate
system for the metric $\eta$ is problematical. A related problem is to relate the Euler vector
field (\ref{euler}) with the vector field
\[
E^{\prime} = \sum_{i=1}^{M+N} u^i \frac{\partial~}{\partial u^i}\,,
\]
the two being equal in the undeformed case.

The various structures on the manifold may be encoded in the deformed (or Dubrovin) connection
\[
{}^D\nabla_XY = \nabla_X Y + z \, X \circ Y\,, \qquad z \in \mathbb{P}^1\,.
\]
For this connection to be torsion free and flat one requires commutativity and associativity of the
multiplication, flatness of the Levi-Civita connection $\nabla$ and potentiality, and visa-versa.
Solutions of the system ${}^D\nabla_\alpha \zeta_\beta$ are automatically gradients,
$\zeta_\alpha=\partial_\alpha {\tilde t}\,.$ Expanding ${\tilde t}=\sum_n \psi^{(n)} z^n$ yields
the recursion relation
\[
\frac{\partial^2 \psi^{(n)}}{\partial t^i \partial t^j} + c_{ij}^k \frac{\partial \psi^{(n-1)}}{\partial t^k}=0\,.
\]
Starting with the seed solutions $\psi^{(0)}=t^i\,,i=1\,,\ldots {\rm dim}\,\mathcal{M}$ one may construct
a fundamental system of solutions.

\subsection{Algebraic deformation theory}

In this section we examine the linearity of the structure
functions of the Frobenius algebra with respect to the parameters $k^i$ from the point of view of
deformation theory (we follow the notation of \cite{book}). Let
\[
M^k(V)=\{m:\underbrace{V\times\ldots \times V}_{k} | m {\rm~linear~in~each~arguement}\}
\]
Recall that a bilinear map $c\in M^2(V)$ defines an associative
structure if and only if
\[
[c,c]_\mathcal{G}=0\,,
\]
where $[\cdot,\cdot]_\mathcal{G}$ is the Gerstenhaber bracket. Owing to
the super-Jacobi identity one has $\delta_c^2=0\,,$ where
\[
\delta_c = [c,\cdot]_\mathcal{G}\quad : M^\bullet(V) \rightarrow M^{\bullet+1}(V)
\]
and this gives rise to the Hochschild complex of $(V,c)\,.$

From proposition \ref{cannotthinkofone} we have the following structure
\[
c(k) = c^{(0)} + \sum_i k_i c^{(i)}\,,
\]
that is, linearity of the structure functions of the associative algebra.
Decomposing the condition $[c(k),c(k)]_\mathcal{G}=0$ for all $k$ one
obtains the following conditions:
\begin{eqnarray*}
[c^{(0)},c^{(0)}]_\mathcal{G} & = & 0 \,,\\
{[c^{(0)},c^{(i)}]}_\mathcal{G} & = & 0 \,,\quad i=1\,,\ldots\,, M\,, \\
{[c^{(i)},c^{(j)}]}_\mathcal{G} & = & 0 \,,\quad i,j=1\,,\ldots\,, M\,.
\end{eqnarray*}
Thus each $c^{(i)}\,, i=0\,,1\,,\ldots\,,M$ separately defines and associative
structure on $T\mathcal{M}\,.$ Each of these define a map $\delta_{c^{(i)}}$
and each $c^{(i)}$ is a cocycle with respect to each cohomology map $\delta_{c^{(j)}}\,,$
that is:
\begin{eqnarray*}
[c^{(i)},c^{(i)}]_\mathcal{G} & = & 0\,,\quad i=0\,,1\,,\ldots\,,M\,,\\
\delta_{c^{(i)}} c^{(j)} & = & 0\,,\quad i,j=0\,,1\,,\ldots\,,M\,.
\end{eqnarray*}
It is also interesting to note that the pair $(\circ,E)$ satisfy the
conditions
\[
\mathcal{L}_{X\circ Y}(\circ) = X \circ \mathcal{L}_Y(\circ)+Y \circ \mathcal{L}_X(\circ)
\]
(following from the semi-simplicity of the multiplication) and the pseudo-scaling
condition
\[
\mathcal{L}^{ext}_E(\circ) = d\,\circ\,.
\]
If one had $\mathcal{L}_E(\circ) = d\,\circ$ then one would have a $F$-manifold \cite{hert}. Here on
has a modified version, where the scaling condition is replaced by the pseudo-scaling
condition. One could also regard the multiplication as defining a deformation of the
$F$-manifold based on the orbit space $\mathbb{C}^N/A_N\,.$

\section{Further Results}

An immediate question these result raise is whether or not the
ideas may be applied to other classes of Frobenius manifolds, the
obvious potential generalization being to other Coxeter orbit
spaces $\mathbb{C}^n/W$, for an arbitrary Coxeter group $W\,.$
By this we mean is there a prepotential schematically of the form
\[
F({\bf t},{\bf b}) = F_W({\bf t}) + k F^{(1)}({\bf t},{\bf b}) + k^2 F^{(2)}({\bf t},{\bf b})
\]
based on the $\mathbb{C}^n/W$ prepotential $F_W$ which is pseudo-quasi-homogeneous
with respect to some suitable Euler field.

For the group $W=B_n$ this is immediate, using the idea originally due to Zuber \cite{zuber},
of embedding the group $B_n$ as a subgroup of $A_{2n+1}\,,$ or geometrically, as the
$B_n$ Frobenius manifold as the induced manifold on certain hyperplanes submanifolds in the
$A_{2n+1}$ Frobenius manifold. This idea generalizes to water-bag type type reductions and this
will be presented in section \ref{BN}.

Another possible generalization, already alluded to above, is to replace the polynomial
part of $\lambda$ by an arbitrary rational function, generalizing the construction
of \cite{kodama, kodama2}. The Frobenius manifold structure on the
space of such rational functions has been much studied and these results can be generalized
to include logarithmic terms. These results are presented in section \ref{rational}

\subsection{$B_N$-type Reductions}\label{BN}

The $B_n$ Frobenius manifold may be regarded as a submanifold in the $A_n$ Frobenius manifold \cite{zuber}.
This idea generalizes to water-bag type potentials.

 \begin{prop}
 On the space of functions
 $$\lambda(p)=p^{2N+2}+s_1p^{2N}+s_3p^{2N-2}+\dots+s_{2N+1}+\sum_{i=1}^{M}k_i\log(p^2-b_i^2)$$
 the formulas (\ref{metric}) and (\ref{multiplication}) define a pseudo-quasi-homogeneous
solution of the WDVV equations.
 \end{prop}
 \begin{proof}
 The function $\lambda$ above is obtained from the following
 waterbag deformation of the $A_{2N+1}$ superpotential:
 \begin{eqnarray*}
 \lambda_A(p)&=&p^{2N+2}+s_1p^{2N}+s_2p^{2N-1}+s_3p^{2N-2}+\dots+s_{2N+1}\\
 &&+\sum_{i=1}^{M}k_i\log(p-b_i) +
 \sum_{i=1}^{M}k_i\log(p-b_{i+M})\,.
 \end{eqnarray*}
 We restrict this to the submanifold
 \begin{eqnarray*}
 s_r&=&0\text{ for }r\text{ even,}\\
 b_{i+M}&=&-b_i\text{ for }1\leq i\leq M\,.
 \end{eqnarray*}
 The restriction of the $s_r$ may be achieved in flat coordinates by setting
 all $t^i$ of odd degree (i.e.\ even $i$) to zero. We introduce
 new flat coordinates $\tilde{b}_i=b_i$ and
 $\tilde{d}_i=b_i+b_{i+M}$ ($i=1,\dots,M$), and restrict to
 $\tilde{d}_i=0$. We check the following components of the
 multiplication tensor restrict to zero on this hyperplane:
 \begin{eqnarray*}
 c_{\tilde{b}_i\tilde{b}_j}^{\tilde{d}_k}\,,&
 c_{\tilde{b}_i\tilde{b}_j}^{t^r}\text{ for }r\text{ even,}\\
 c_{\tilde{b}_i t^r}^{\tilde{d}_k}\text{ for }r\text{ odd,} &
 c_{\tilde{b}_i t^r}^{t^s}\text{ for }r\text{ odd, }s\text{ even,}\\
 c_{t^r t^s}^{\tilde{d}_k}\text{ for }r,s\text{ odd,} &
 c_{t^r t^s}^{t^u}\text{ for }r,s\text{ odd, }u\text{ even.}
 \end{eqnarray*}
 Polynomial terms arising in these components can be seen to
 vanish from consideration of their degrees; all polynomials in
 $\{t^1,\dots,t^{2N+1}\}$ of odd degree must vanish when all $t^i$ of odd degree vanish,
 whereas polynomials in $\{t^i\}$ of even degree are
 always multiplied by (at least) a factor of $b_i+b_{i+M}$ for
 some $i$, and hence vanish on $d_i=0$. Non-polynomial terms are
 given explicitly in Lemma \ref{lemma:multiplication}.

 \end{proof}

 \noindent It would be of interest to see if these ideas can be applied to arbitrary
 Coxeter group orbit space.

 \subsection{Rational Water-bag Potentials}\label{rational}
 \begin{prop}
 On the space of functions
 \begin{eqnarray*}
 \lambda(p)&=&p^{N+1}+s_1p^{N-1}+\dots+s_N\\
 &&+\sum_{i=1}^{K}\left[\frac{v_{i,1}}{(p-s_i)}+\dots+\frac{v_{i,L_i}}{(p-s_i)^{L_i}}\right]\nonumber\\
 &&+\sum_{i=1}^M k_i\log(p-b_i)\nonumber\,,
 \end{eqnarray*}
 the formulas (\ref{metric}) and (\ref{multiplication}) define a
 solution of the WDVV equations.
 \end{prop}
 \begin{proof}
 Canonical coordinates are found as in Lemma \ref{lem:compat}.

 The flat coordinates are $\{b_1,\dots,b_M\}$ together with those
 obtained for the purely rational case \cite{kodama},\cite{aratyn}. Namely invert
 $\lambda_+(p)=p^{N+1}+s_1p^{N-1}+\dots+s_N$ about $p=\infty$
 using the Puiseaux series (\ref{puiseaux}), and invert
 $$\lambda_{-i}(p)=\frac{v_{i,1}}{(p-s_i)}+\dots+\frac{v_{L_i}}{(p-s_i)^{L_i}}$$ for $p\sim
 s_i$ as
 $$p=\frac{1}{L_i}\left(x_{i,L_i+1}+\frac{x_{i,L_i}}{w}+\dots+\frac{x_{i,1}}{w^{L_i}}\right)\,,$$
 where $\lambda_{-i}=w^{L_i}$, and $x_{i,L_i+1}=L_is$. The flat coordinates are then
 $\{t^\alpha, x_{\beta,\gamma}, b_\delta\}$.  In these coordinates the metric
 has only the following non-zero components:
 \begin{eqnarray*}
 \eta\left(\dbyd{t^\alpha},\dbyd{t^\beta}\right)&=&-\frac{1}{N+1}\delta_{\alpha+\beta,N+1}\,,\\
 \eta\left(\dbyd{x_{i,j}},\dbyd{x_{i,k}}\right)&=&-\frac{1}{L_i}\delta_{j+k,L_i+2}\,,\\
 \eta\left(\dbyd{b_\alpha},\dbyd{b_\beta}\right)&=&k_\alpha\delta_{\alpha\beta}\,.
 \end{eqnarray*}
 \end{proof}

 \noindent Note one may combine the results from the last to sections and consider
 $B_n$-type reductions of the rational case, where the superpotential, including
 logarithmic terms, is an even function.

 \medskip

In the above proposition the location of the poles $\{s_i\}$ and the logarithmic
poles $\{b^i\}$ were taken to be distinct. However, a modified of the above proposition
may be formulated which takes into account possible coincidences in these
sets. Rather than state this we give an example.

\begin{example} The superpotential
\[
\lambda(p)  = p^2  + t_1 + \frac{t_2}{(p-t_3)} + k \log(p-t_3)
\]
leads to the following solution of the WDVV equation
\begin{equation}
F= \frac{1}{12} t_1^3 +t_1 t_2 t_3 - \frac{1}{2} k \, t_1 t_3^2 - \frac{3}{4} t_2^2 + \frac{1}{2} t_2^2 \log t_2  +
\frac{1}{3} t_2 t_3^3 - \frac{1}{12} k \,t_3^4\,.
\label{eawexample}
\end{equation}
\end{example}

\noindent This produces an interesting class of solutions, as no extra variables have had to be
introduced, so in a sense they are true deformations of the original solution. The single pole
case - generalizations of the above example - are isomorphic to deformations of the
extended-affine-Weyl orbit space \cite{DZ}, since
\[
H_{0,N+L+1}(N+1,L) \cong \mathbb{C}^{N+L+1}/{{\widetilde{W}}}^{(L)}(A_{N+L})\,.
\]
Explicitly this is given by a Legendre transformation (which acts on solutions of the WDVV equations, not
just to those solutions which define Frobenius manifolds).
\begin{example}
Applying the Legendre transformation $S_2$ (using the notation of \cite{dubrovin1}) to the
solution (\ref{eawexample}) yields the solution
\[
{\hat F} = \frac{1}{4} {\hat t}_1 + \frac{1}{2} {\hat t_2}^2 {\hat t}_3 - \frac{1}{2} k \,{\hat t}_2 {\hat t}_3^2 -
\frac{1}{96}{\hat t}_1^4 + {\hat t}_1 e^{{\hat t}_3} -
k
\left( \frac{1}{4} {\hat t}_1^2 {\hat t}_3 + \frac{1}{2}{\hat t}_2 {\hat t}_3^2 \right)+
\frac{1}{6} k^2 {\hat t}_3^3\,.
\]
This defines a deformation of the extended-affine-Weyl space $\mathbb{C}^3/{{\widetilde{W}}}^{1}(A_2)\,.$
\end{example}
\noindent One would expect that the associated dispersionless integrable systems would be related to
water-bag type-reductions of the dispersionless Toda equations and their generalizations \cite{BogKon}.

\section{Open problems}

Some open problems have already been outlined above; here we draw them together
and raise some other open problems, potential generalizations and applications.

\begin{itemize}

\item{} Can the construction be applied, independent of the Landau/Ginzburg
construction, directly to an arbitrary Coxeter group orbit space, or more generally,
to other orbit spaces? By this we mean, is there a Saito-type construction
of these solutions? The absence of a flat \lq intersection form\rq~would
seem problematical. A related question is whether one can formulate axiomatically
a theory of pseudo-quasi-homogeneous solutions of the WDVV equations.

\item{} The Frobenius manifold structure on the space of rational functions
may be generalized to the space of branched coverings of an arbitrary Riemann
surface (i.e. a Hurwitz space). All that is required for the direct calculation
of the residues (\ref{metric}) and (\ref{multiplication}) is the meromorphicity of the
{\sl derivatives} of $\lambda$ rather than the meromorphicity of $\lambda$ itself. This
suggests that one should look at generalizations where $\lambda$ lies in some
extension of the field of meromorphic functions.

\item{} In a semi-simple Frobenius manifold there exists interesting submanifolds:
discriminants and caustics \cite{iabs}. What are the properties of such structures in the
present case?

\item{} What are the properties of the dispersionless integrable systems associated to
such solutions of the WDVV equations, i.e. the water-bag reductions of the dKP hierarchy
itself, and how are they encoded in the geometry of these pseudo-quasi-homogeneous
manifolds? In particular, the (non-local) bi-Hamiltonian structure, especially in the flat coordinates
system for the metric $\eta$ is unknown in general. Can these dispersionless systems
be deformed, and how do the form of such deformations follow from the geoemtry of the
undeformed systems \cite{chang}.

\item{} Finally, is there an algebraic description, say of the $A_n$-deformations, in terms
of a deformed Milnor ring? Are there field theoretic interpretation of the results in terms of
a topological quantum field theory \cite{k1, k2}.

\end{itemize}

\noindent We hope to address some of these problems in the future.

\section*{Acknowledgments} James Ferguson would like to thank the Carnegie Trust for the Universities
of Scotland for a research studentship.


\begin{thebibliography}{99}


\bibitem{kodama} Aoyama, S. and Kodama, Y., {\sl Topological conformal field theory with a rational W
potential and the dispersionless KP hierarchy},Mod. Phys. Lett.
{\bf A 9} (1994) 2481-2492.

\bibitem{kodama2} Aoyama, S. and Kodama, Y., {\sl Topological Landau-Ginzburg with a rational
potential, and the dispersionless KP hierarchy}, Comm. Math. Phys. {\bf 182} (1996) 185-219.


\bibitem{BogKon} Bogdanov, L.V. and Konopelchenko, B.G., {\sl Symmetry constraints for
dispersionless integrable equations and systems of hydrodynamic type}, Phys. Lett. {\bf A 330}
(2004) 448-459.

\bibitem{book} Cannas da Silva, A. and Weinstein, A., {\sl Geometric Models for Noncommutative Algebras},
Berkeley Mathematics Lecture Notes, Vol. 10 (A.M.S. publications, 1999).


\bibitem{chang} Chang, J-H., {\sl On the water-bag model of dispersionless KP hierarchy},
arXiv:nlin.SI/0603007.

\bibitem{DS} David, L., and Strachan, I.A.B., {\em Compatible metrics
on manifolds and non-local bi-hamiltonian structures}, Int.
Math. Res. Notices, {\bf 66} (2004), 3533-3557.

\bibitem{dubrovin1} {Dubrovin, B., {\sl Geometry of 2D topological field theories} in {\sl Integrable
Systems and Quantum Groups}, ed. Francaviglia, M. and Greco, S..
Springer lecture notes in mathematics, {\bf 1620}, 120-348.}

\bibitem{dubrovin2} Dubrovin, B., {\em On almost duality for Frobenius manifolds} in
{\sl Geometry, topology, and mathematical physics}, 75--132, Amer.
Math. Soc. Transl. Ser. 2, 212, Amer. Math. Soc., Providence, RI,
2004.

\bibitem{DZ} Dubrovin, B. and Zhang, Y., {\em Extended affine Weyl groups and Frobenius manifolds},
Compositio Mathematica {\bf 111} (1998), 167--219.

\bibitem{FV} Feigin, M.V. and Veselov, A.P., {\sl Coxeter discriminants and logarithmic Frobenius structures},
math-ph/0512095.

\bibitem{GibbonsT} Gibbons, J. and Tsarev, S.P., {\sl Reductions of the Benney Equations}, Phys. Lett.
{\bf A 211} (1996) 19-24.

\bibitem{hert} Hertling, C.,
{\em Frobenius manifolds and moduli spaces for singularities},
Cambridge Tracts in Mathematics, {\bf 151}, Cambridge University
Press (2002).

\bibitem{k1} Krichever, I.M., {\sl The dispersionless equations and topological minimal models},
Comm. Math. Phys. {\bf 143(2)} (1992), 415-429.

\bibitem{k2} Krichever, I.M., {\sl The $\tau$-function of the universal Whitham hierarchy, matrix models,
and topological field theories}, Comm. Pure Appl. Math {\bf 47} (1994) 437-475.

\bibitem{MH} Martini, R., and Hoevenaars, L.K., {\sl Trigonometric Solutions of the WDVV Equations from Root Systems},
Lett.Math.Phys. {\bf 65} (2003) 15-18.

\bibitem{pavlov} Pavlov, M.V., {\sl The Hamiltonian approach in classification and
integrability of hydrodynamic chains}, arXiv:nlin.SI/0603057


\bibitem{iabs} Strachan, I.A.B., {\sl Frobenius manifolds: natural submanifolds and induced bi-Hamiltonian
structures}, Differential Geom. Appl. {\bf 20} (2004) 67-99.


\bibitem{saito} Saito, K., {\sl On a linear structure of a quotient variety by a finite reflection group}\,,
Preprint RIMS-288 (1979).



\bibitem{zuber} Zuber, J.-B., {\sl On Dubrovin Topological Field
Theories} Mod. Phys. Lett. {\bf A 9} (1994) 749-760.


\end{thebibliography}
\end{document}